\documentclass[intlimits,twoside,a4paper]{article}

\usepackage[cp1251]{inputenc}
\usepackage[eqsecnum]{cmpj3}
\issue{2022}{25}{3}{33701}
\doinumber{10.5488/CMP.25.33701}

\title[Ferromagnetic Heisenberg model]%
{Ferromagnetic Heisenberg model with the Dzyaloshinskii-Moriya interaction%
}
\author[E. Albayrak]{E. Albayrak\orcid{0000-0003-2695-0912}\thanks{Corresponding author: \email{albayrak@erciyes.edu.tr}.}}
\address{ Physics Department, Erciyes University, 38039 Kayseri, T\"{U}RK\.{I}YE
}
\date{Received May 30, 2022, in final form July 25, 2022}

\begin{document}
	
\maketitle
\begin{abstract}
The spin-1/2 Heisenberg model is formulated in terms of a mean-field approximation (MFA) by using the matrix forms of spin operators $\hat{S}_x,\hat{S}_y$ and $\hat{S}_z$ in three-dimensions. The considered Hamiltonian consists of bilinear exchange interaction parameters $(J_x,J_y,J_z)$, Dzyaloshinskii-Moriya interactions $(\Delta_x,\Delta_y,\Delta_z)$ and external magnetic field components $(H_x,H_y,H_z)$. The magnetization and its components are obtained in the MFA with the general anisotropic case with $J_x\neq J_y \neq J_z$ for various values of coordination numbers $q$. Then, the thermal variations of magnetizations 
are investigated in detail to obtain the phase diagrams of the model for the isotropic case with $J_x=J_y=J_z>0$. It is found that the model exhibits ferromagnetic, paramagnetic, random phase regions and an extra ferromagnetic phase at which the components of magnetizations present branching.
%
%
%\keywords Up to six keywords (\href{https://physh.aps.org/browse}{Physics Subject Headings})
\printkeywords
%
%\pacs 75.10.Hk;75.30.Kz;75.50.Gg
\end{abstract}

\section{Introduction}
%\doclicenseThis
The Dzyaloshinskii-Moriya (DM) interaction~\cite{DM, DM2} is an antisymmetric exchange interaction with a contribution to total magnetic exchange interaction between two neighboring magnetic spins as a source of weak ferromagnetic (FM) behavior in an antiferromagnet~\cite{Komatsu}. The helical structure is induced by the DM interaction which breaks the chiral symmetry, and thus, the two same helical structures with different winding directions do not degenerate~\cite{Nishikawa}. It is revealed theoretically that a chiral magnetic soliton lattice is formed with a finite magnetic field perpendicular to the axis of the helical structure, and a continuous phase transition to a forced ferromagnetic phase occurs with increasing magnetic field~\cite{Shinozaki}. These topologically protected magnetic structures are called skyrmions and their existence was proven experimentally~\cite{ExpSky, ExpSky2, ExpSky3, ExpSky4, ExpSky5, ExpSky6, ExpSky7, ExpSky8}. Due to its potential applications in spintronics devices, they have drawn much attention.  Note also that the competition between DM and exchange interactions is important for skyrmion-lattice phases. Even though the DM interaction has been well studied in crystalline magnets, there are also numerous experimental~\cite{ExpSG, ExpSG2} and theoretical~\cite{TheoSG, TheoSG2} studies demonstrating its importance in spin glasses. 

The DM interaction may be rather difficult to deal with analytically and thus it may require numerical investigations. Because of the nature of DM interaction, it is necessary to study it with the spins at least in two-dimensions in contrast to the Ising spins. The quantum phase transitions (QPT) in a bond-alternative antiferromagnetic (AFM) Ising chain were considered including the DM interaction~\cite{BLi} by using the fidelity based on the infinite matrix product states algorithm. It was found that antiferromagnetic and disordered phases exist in the ground state and the transition between them is continuous. The low-energy excitation spectrum and the ground-state magnetic phase diagram of the spin-1/2 ferromagnetic Ising chain with the DM interaction were considered where a first-order metamagnetic phase transition between FM and a spiral phases were reported~\cite{Soltanix}. 

The spin-1/2 models with the inclusion of the DM interaction (DMI) were also established in some works. The Ising--Heisenberg model on the triangulated kagome lattice was exactly solved by establishing a precise mapping correspondence to the simple spin-1/2 Ising model~\cite{Strecka}. The quantum anisotropic Heisenberg antiferromagnet model in the presence of a DMI and a uniform longitudinal magnetic field was considered  by using the effective-field theory with a finite cluster $N=2$ spin~\cite{Parente}. The QPTs and ground-state phase diagram of the Heisenberg–Ising alternating chain with uniform DM interaction were investigated by a matrix-product-state method~\cite{Liu1}. The pair XYZ Heisenberg interaction and quartic Ising interactions were exactly solved by establishing a precise mapping relationship with the corresponding zero-field eight-vertex model~\cite{Strecka1}. The anisotropic Heisenberg model with antiferromagnetic exchange interactions in the presence of a longitudinal external magnetic field and a DMI was studied by employing the usual mean-field approximation~\cite{Parente1}. The ground-state magnetic phase diagram of an antiferromagnetic two-leg ladder with period two lattice units modulated DMI along the legs was considered~\cite{Avalishvili}. The results of a combined analytical and density matrix renormalized group study of the AFM XXZ Heisenberg chain subject to a uniform DMI and a transverse magnetic field were reported~\cite{Chan}. The linked cluster expansion techniques was applied to study the polarized high-field phase of an antiferromagnet on the kagome lattice with Heisenberg and DMI~\cite{Flynn}. The phase diagram in the $H$-$T$ plane of the potassium jarosite compound KFe$_{3}$(OH)$_{6}$(SO$_{4}$)$_{2}$ for the antiferromagnetic XY model with DMI was investigated by using the mean-field theory for different values of DM~\cite{Freitas}. On the two-dimensional non-linear $\Sigma$-model describing a ferromagnet with DMI, three families of exact static solutions depending on a single Cartesian variable were obtained~\cite{Grandi}. The roles of spatial anisotropy, DM interactions and quantum fluctuations on the magnetization process of a triangular antiferromagnet were considered~\cite{Griset}. The ground-state phase diagram of a one-dimensional XXZ chain with a spatially modulated DMI in the presence of an alternating magnetic field was calculated~\cite{Japaridze}. The steady-state phase diagram of the dissipative XYZ model on a two-dimensional triangular lattice by means of cluster mean-field approximation was constructed~\cite{XLi}. The last work that we can mention is the study of the quantum spin liquid material herbertsmithite described by an AFM Heisenberg model on the kagome lattice~\cite{Messio}.

In addition to the above theoretical works, we can also mention some experimental works such as: magnetization of triangular-lattice antiferromagnets Ba$_3$CoSb$_2$O$_9$ and CsCuCl$_3$ with a $120^{\rm o}$ spin structure in the $ab$ plane~\cite{Sera}, the nature of possible magnetic phases in the frustrated hyperkagome iridate, Na$_4$Ir$_3$O$_8$, based on the Kitaev-Heisenberg model with DM interactions~\cite{Shindou}; the results of magnetic measurements performed on geometrically frustrated Ni$_3$V$_2$O$_8$ and Ni$_{3(1-x)}$Co$_{3x}$V$_2$O$_8$ single crystals with $x = 0.03$ were presented~\cite{Szymczak}.

In this work, the magnetization components $(M_x,M_y,M_z)$ and thus magnetization ($M_T$) are obtained in terms of the MFA for the general values of the $(J_x,J_y,J_z)$, $(\Delta_x,\Delta_y,\Delta_z)$ and $(H_x,H_y,H_z)$ and $q$. Then, thermal variations of magnetizations are studied for ferromagnetic interactions between the nearest-neighbor spins, i.e., $J=J_x=J_y=J_z>0$, with equal DM interactions and external magnetic field components. In addition, the phase diagrams are obtained on the planes of ($\Delta_m/J, T/J$) for $q=3, 4$ and 6 when $H$ is set equal to zero. Even though our equations are obtained for the anisotropic case, the results are only presented for the isotropic case for simplicity. The reason of this study is that our literature search did not even reveal any results for the three-dimensional isotropic case.  

The rest of the work is set up as follows: the formulation for the MFA in terms of the spin operators is presented in section 2, thermal variations of magnetizations are demonstrated in section 3, phase diagrams are illustrated in section 4 and the last section includes a brief summary and conclusions.

\section{The formulation}
The anisotropic spin-1/2 XYZ Heisenberg Hamiltonian in terms of the bilinear exchange interaction parameters and the DM interactions between the nearest-neighbor (NN) spins and the external magnetic field components acting on each spin site is given as
\begin{eqnarray} \label{eq:Hamiltonian}
	\mathcal{H}&=&-J_x \sum_{\langle i,j \rangle} S_{i}^x S_{j}^x-J_y \sum_{\langle i,j \rangle} S_{i}^y S_{j}^y-J_z \sum_{\langle i,j \rangle} S_{i}^z S_{j}^z -\Delta_x \sum_{\langle i,j \rangle} (S_{i}^y S_{j}^z-S_{i}^z S_{j}^y) -\Delta_y \sum_{\langle i,j \rangle} (S_{i}^z S_{j}^x-S_{i}^x S_{j}^z)   \nonumber \\
	&-&\Delta_z \sum_{\langle i,j \rangle} (S_{i}^x S_{j}^y-S_{i}^y S_{j}^x)-H_x \sum_{i}  S_{i}^x-H_y \sum_{i}  S_{i}^y-H_z \sum_{i}  S_{i}^z,
\end{eqnarray}
where $\langle i,j \rangle$ refers to the summation over the NN spins. $\hat{S}_{i}^x, \hat{S}_{i}^y$ and $\hat{S}_{i}^z$ are the components of spin-1/2 operator at site $i$ which are given in the matrix forms as
\begin{eqnarray}\label{eq:SO}
	\hat{S}_{i}^x=\frac{1}{2}
	\left(\begin{array}{cc}
		0 & 1   \\
		1 & 0   \\
	\end{array}\right), \hspace{0.05cm} 
	\hat{S}_{i}^y=\frac{1}{2}
	\left( \begin{array}{cc}
		0 & -i  \\
		i & 0   \\
	\end{array}\right) \hspace{0.05cm}, \hspace{0.05cm}
	\hat{S}_{i}^z=\frac{1}{2}
	\left(\begin{array}{cc}
		1 & 0  \\
		0 & -1 \\
	\end{array}\right).
\end{eqnarray}
In the MFA, the Hamiltonian $\mathcal{H}$ in equation~(\ref{eq:Hamiltonian}) can be written in the MF form as
\begin{eqnarray} \label{eq:MFA}
	-\beta \mathcal{H}_{\rm {MFA}}=-\beta \sum_{i} \mathcal{H}_{\rm {MFA}}^{(i)},
\end{eqnarray}
in which
\begin{eqnarray} \label{eq:MFA1}
	-\beta  \mathcal{H}_{\rm {MFA}}^{(i)}&=&\beta q (J_x M_x S_{i}^x+J_y M_y S_{i}^y+J_z M_z S_{i}^z) +\beta (H_x S_{i}^x+H_y S_{i}^y+H_z S_{i}^z) \nonumber \\ 
	&+&\Delta_x (S_{i}^y M_z-S_{i}^z M_y)+\beta \Delta_y (S_{i}^z M_x-S_{i}^x M_z)+\beta \Delta_z  (S_{i}^x M_y-S_{i}^y M_x),
\end{eqnarray}
where $M_{\mu}=\langle S_{j}^{\mu}\rangle$ are the magnetization components with ${\mu}=x, y, z$ and $\beta=1/({\rm k} T)$ with ${\rm k}$ being the Boltzmann constant set equal to 1 for convenience.

The matrix representation of $-\beta \mathcal{H}_{\rm {MFA}}^{(i)}$ is obtained by using the spin operators, i.e., equation~(\ref{eq:SO}), and is found as 
\begin{eqnarray} \label{eq:ParMat}
	-\beta \mathcal{H}_{\rm {MFA}}^{(i)}=
	\left(\begin{array}{cc}
		H_{11}  & H_{12} \\
		H_{21}\ & H_{22} \\
	\end{array}\right),
\end{eqnarray}
where the matix elements are given as  
\begin{eqnarray} H_{11}&=&{\beta}/{2}[q(\Delta_y M_x-\Delta_x M_y+J_z M_z)+H_z], \nonumber \\  
H_{12}&=&{\beta}/{2} [q(J_x M_x+ \Delta_z (\ri M_x + M_y)- \ri J_y M_y- \ri \Delta_x M_z- \Delta_y M_z)+(H_x-\ri H_y)], \nonumber \\
H_{21}&=&{\beta}/{2} [q (J_x M_x + \Delta_z (-\ri M_x+ M_y) + \ri J_y M_y+ \ri \Delta_x M_z- \Delta_y M_z)+(H_x+\ri H_y), \nonumber \\
H_{22}&=&{\beta}/{2} [q (-\Delta_y M_x+\Delta_x M_y-J_z M_z)-H_z]. \nonumber
\end{eqnarray}
It is clear that $H_{12}=H_{21}^*$, which ensures that the eigenvalues of this matrix are real. Thus, the eigenvalues are given as  
\begin{eqnarray}
\Lambda_{1,2}&=& \pm {\beta}/{2} \left[ \left(H_x^2 + H_y^2 + H_z^2 \right) + 2q \left(-\Delta_z H_y M_x + \Delta_y H_z M_x  + H_x J_x M_x  + \Delta_z H_x M_y \right. \right. \nonumber \\ 
&-& \left. \Delta_x H_z M_y 
+ H_y J_y M_y - \Delta_y H_x M_z  + \Delta_x H_y M_z  + H_z J_z M_z \right) +q^2 \left( \Delta_y^2 M_x^2 + \Delta_z^2 M_x^2  \right.  \nonumber \\
&+& J_x^2 M_x^2 - 2 \Delta_x \Delta_y M_x M_y  
+ 2 \Delta_z J_x M_x M_y  - 2 \Delta_z J_y M_x M_y  + \Delta_x^2 M_y^2  + \Delta_z^2 M_y^2  + J_y^2 M_y^2 
\nonumber \\
&-& 2 \Delta_x \Delta_z M_x M_z - 2 \Delta_y J_x M_x M_z  + 2 \Delta_y J_z M_x M_z - 2 \Delta_y \Delta_z M_y M_z  
+ 2 \Delta_x J_y M_y M_z \nonumber \\
&-& \left. \left. 2 \Delta_x J_z M_y M_z + \Delta_x^2 M_z^2 
+ \Delta_y^2 M_z^2  + J_z^2 M_z^2 \right) \right]^{1/2}. \nonumber
\end{eqnarray}
Having obtained the eigenvalues ($\Lambda_1, \Lambda_2$), we are now ready to obtain the partition function which is given as 
\begin{eqnarray} \label{eq:Partition}
	Z_i&=&\mathrm(Tr)_{(i)} \exp[-\beta \mathcal{H}_{\rm {MFA}}^{(i)}]  \nonumber \\
	&=& \sum_{n=1}^2 \re^{\Lambda_n}=\re^{\Lambda_1}+\re^{\Lambda_2}=2 \cosh(\Lambda).
\end{eqnarray}
since $\Lambda=\Lambda_1=-\Lambda_2$. The free energy of the model is found from the well-known definition by using the partition function as
\begin{eqnarray}\label{eq:FreeE}
	f=-\frac{1}{\beta} \ln Z_i
\end{eqnarray}
which will be utilized to find the formulations of the order-parameters.
The dipolar moments or magnetization components, $M_{\mu}=\langle S_{j}^{\mu}\rangle$ with ${\mu}=x, y, z$ are found from
\begin{eqnarray}\label{eq:Sz}
	M_{\mu}&=&\langle S_i^\mu\rangle=-\frac{\partial f}{\partial H_\mu}=\frac{1}{\beta} \frac{\partial \ln Z_i}{\partial H_\mu}=\frac{\mathrm(Tr)_{(i)} \left\{S_i^{\mu} \exp \left[-\beta \mathcal{H}_{\rm {MFA}}^{(i)} \right] \right\}}{Z_i} \nonumber \\
	&=& \frac{1}{\beta} \left[\frac{({\partial \Lambda_1}/{\partial H_\mu})\re^{\Lambda_1}+({\partial \Lambda_2}/{\partial H_\mu})\re^{\Lambda_2}}{\re^{\Lambda_1}+\re^{\Lambda_2}}\right]\nonumber \\&=& \frac{1}{\beta'} \left[\frac{({\partial \Lambda_1}/{\partial h_\mu})\re^{\Lambda_1}+({\partial \Lambda_2}/{\partial h_\mu})\re^{\Lambda_2}}{\re^{\Lambda_1}+\re^{\Lambda_2}}\right],
\end{eqnarray}
where $\beta'=\beta J_z$ and $h_\mu=H_\mu/J_z$ which are the reduced temperature and external magnetic field components. The explicit forms of the magnetization components are too long to be given explicitly. The magnitude of the magnetization vector, $\vec{M}_T$= $M_x\hat{i}+M_y \hat{j}+M_z\hat{k}$, is obtained by using the  magnetization components given in equation~(\ref{eq:Sz}) as 
\begin{eqnarray}
	M_T=\sqrt{\sum_{\mu=x,y,z} M_\mu^2}=\sqrt{M_x^2+M_y^2+M_z^2}.
\end{eqnarray}
Having obtained the formulations for the magnetization components and magnetization in the MFA for the general case, we are now ready to study their thermal variations for the given values of $\Delta_{\mu}$, $J_{\mu}$, $H_{\mu}$ and the coordination number $q=3, 4$ and 6 corresponding to the honeycomb, square and simple cubic lattices. An iterative procedure is followed for the calculation of our order-parameters with the given values of the system parameters under temperature variations. The obtained results are only presented for the FM case with $J=J_x=J_y=J_z>0$ corresponding to the isotropic case, as seen in the next section.

\begin{figure}
	\begin{center}
		\includegraphics[width=6.9cm]{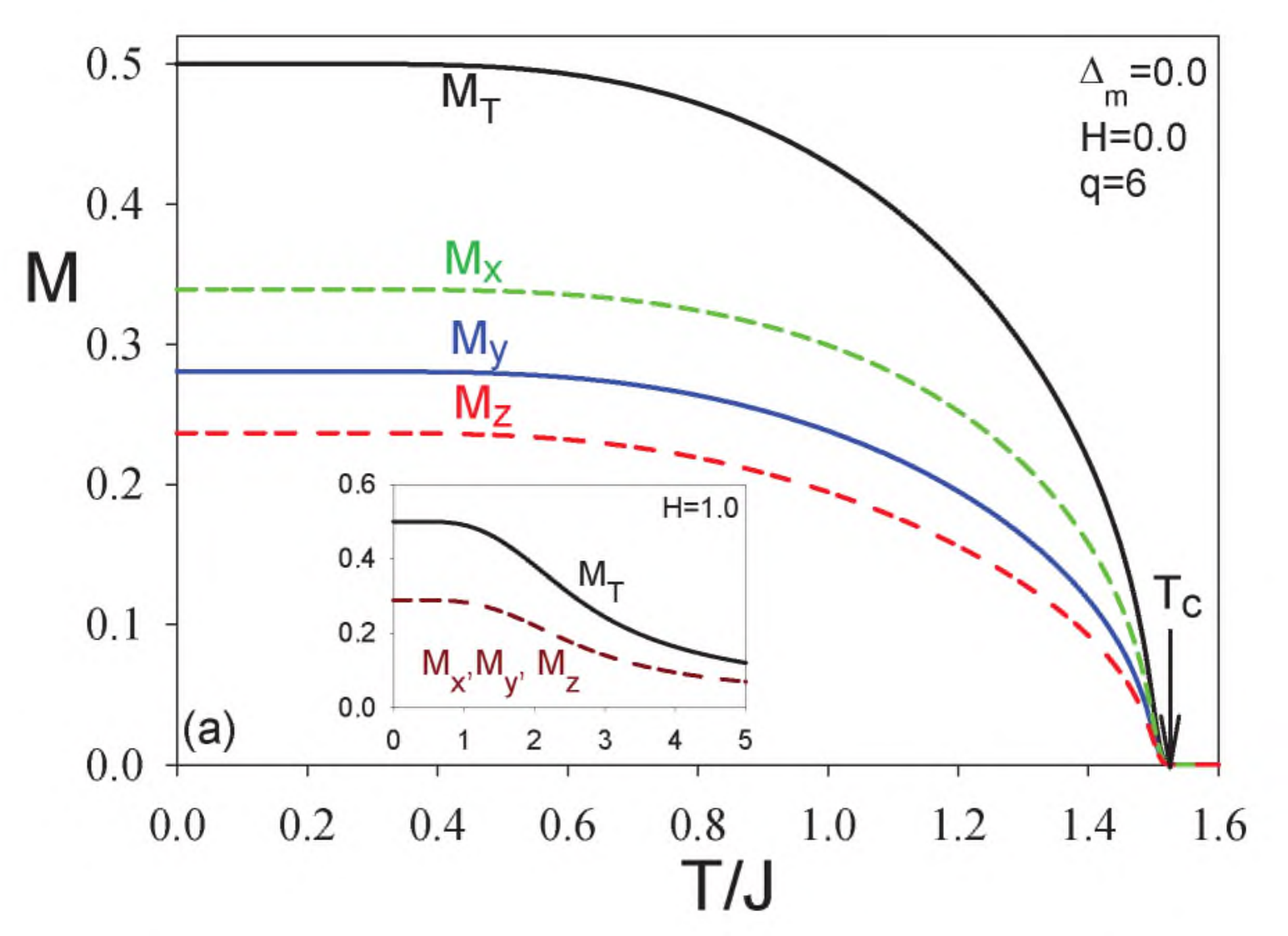}
		\includegraphics[width=6.9cm]{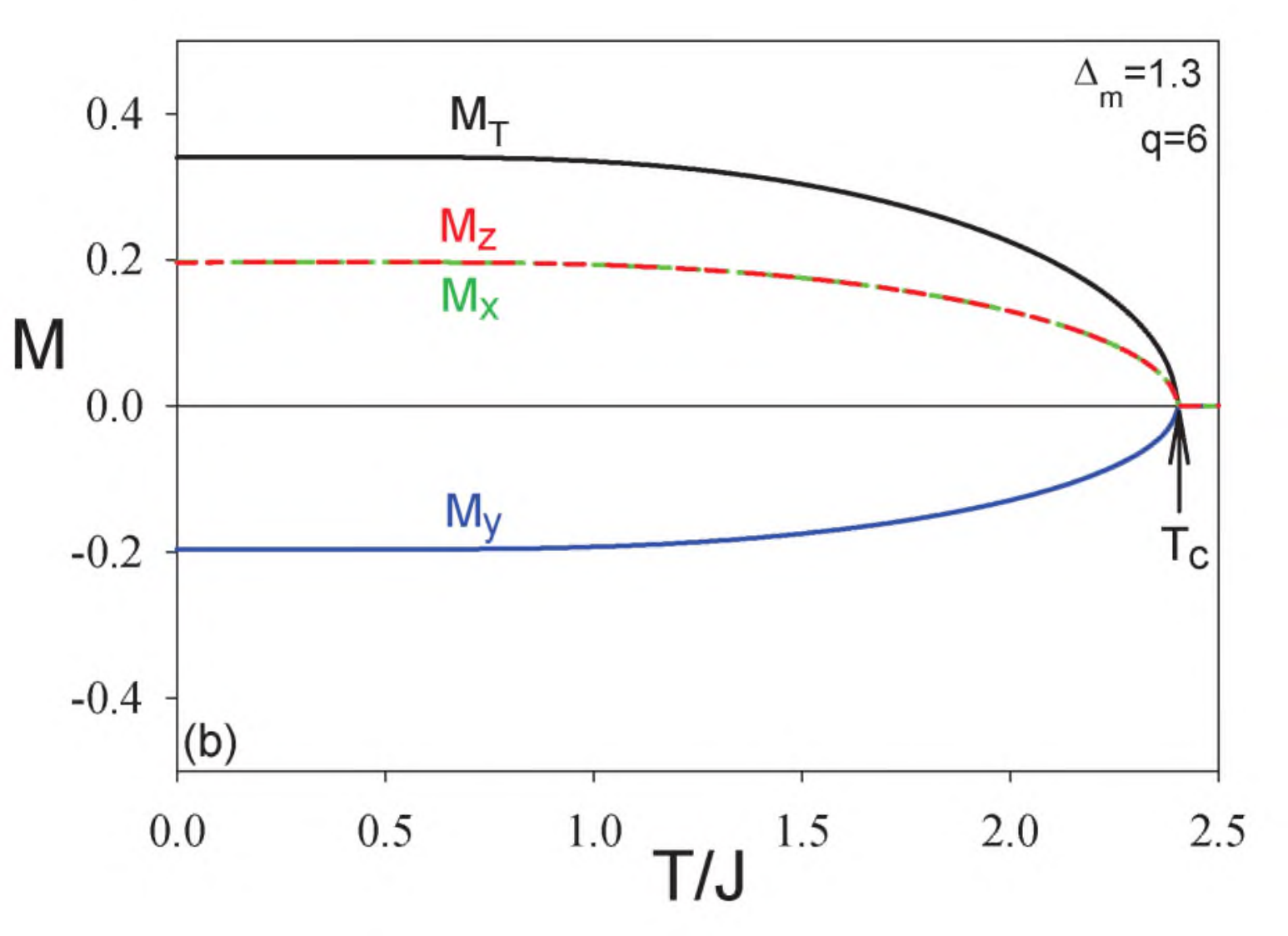}
		\includegraphics[width=6.9cm]{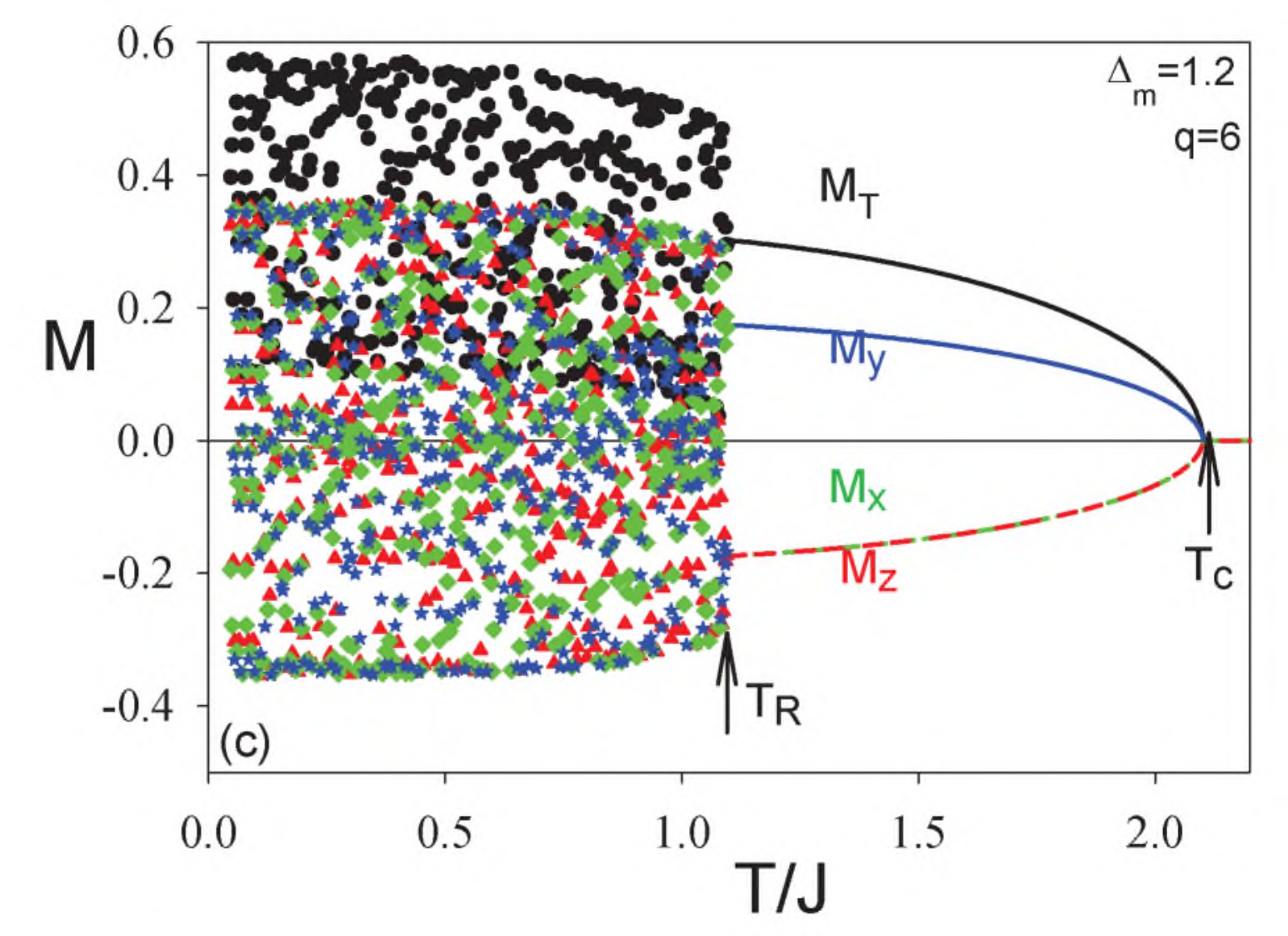}
		\includegraphics[width=6.9cm]{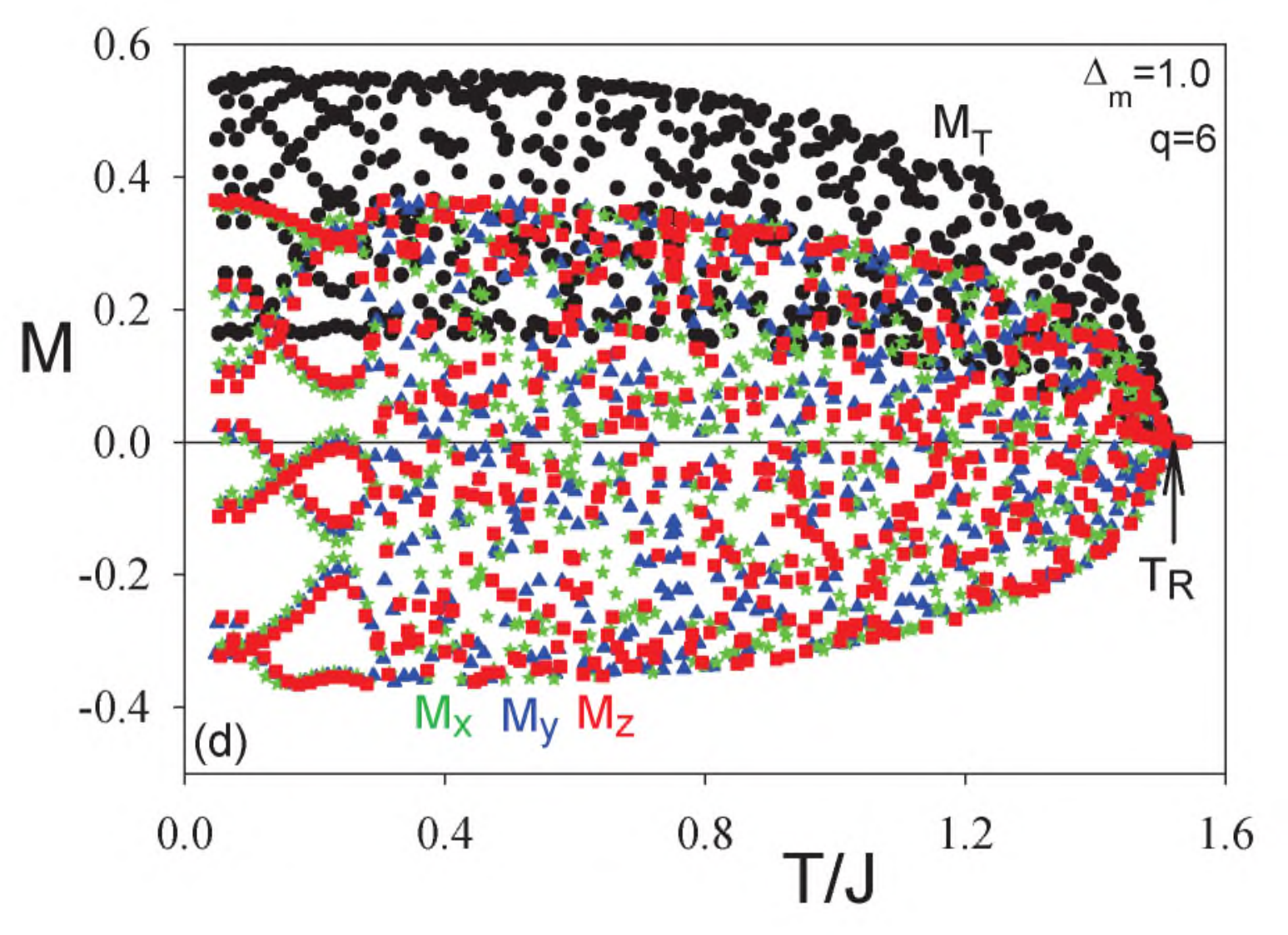}
		\includegraphics[width=6.9cm]{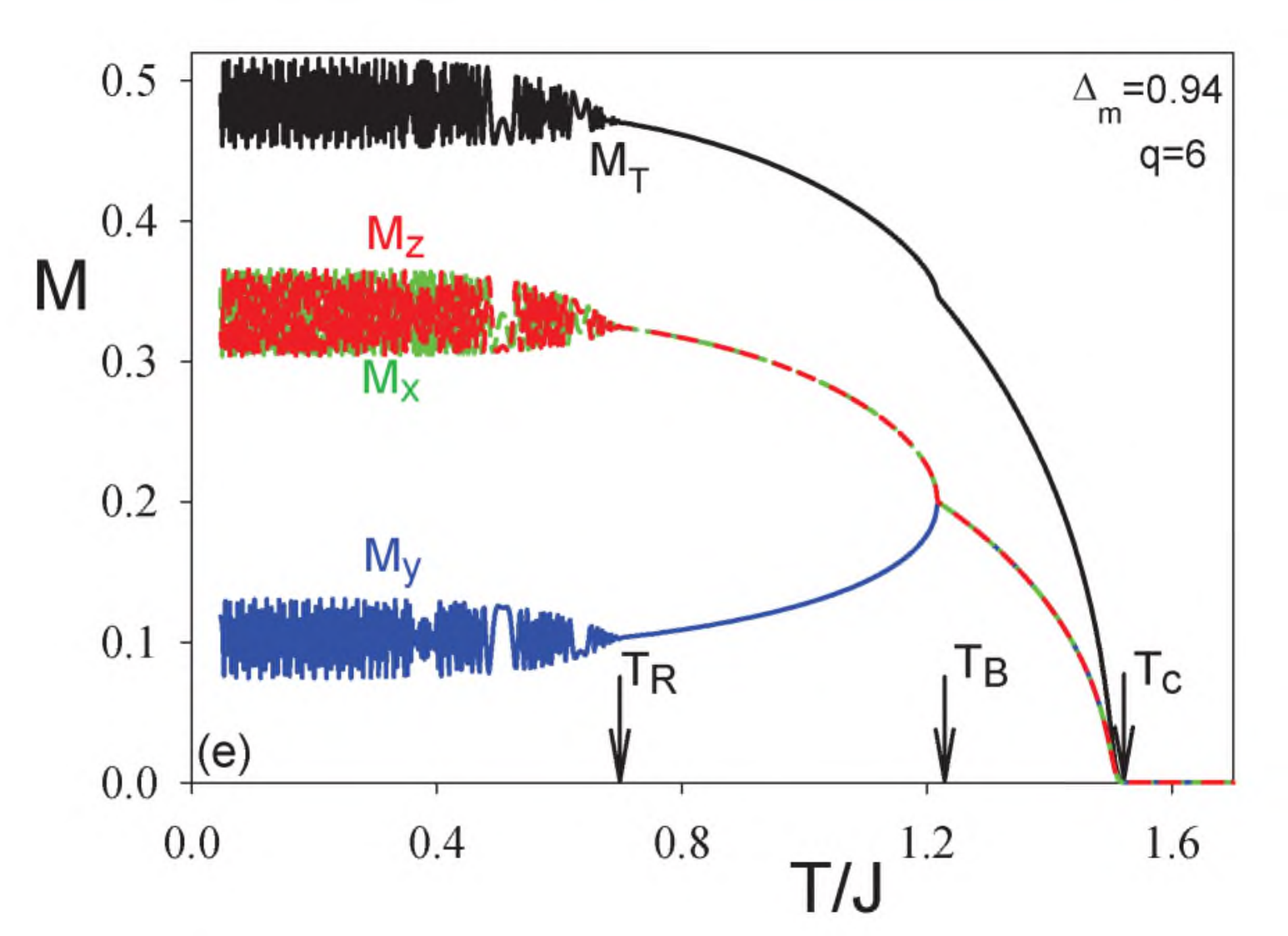}
		\includegraphics[width=6.9cm]{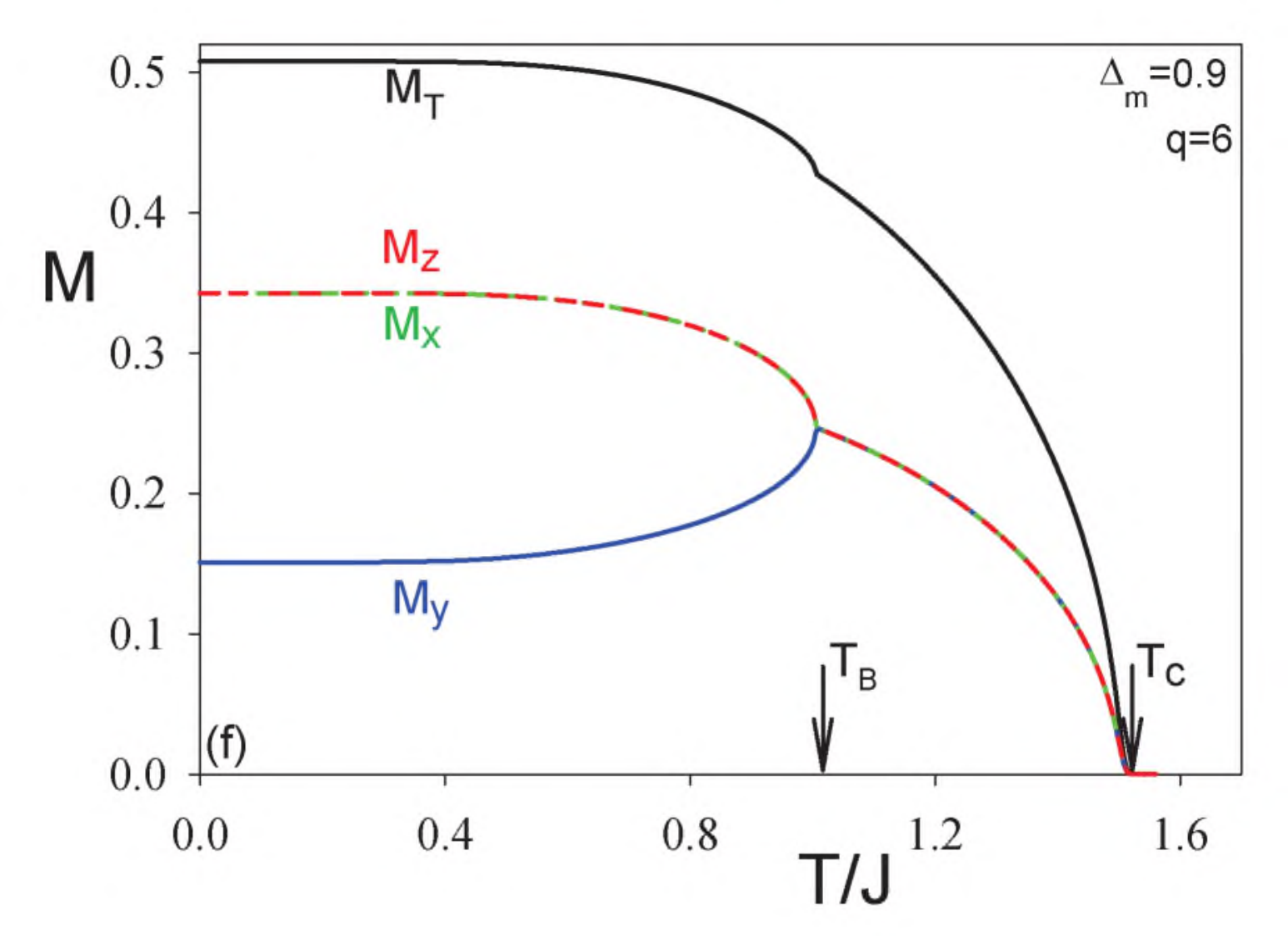}
		\includegraphics[width=6.9cm]{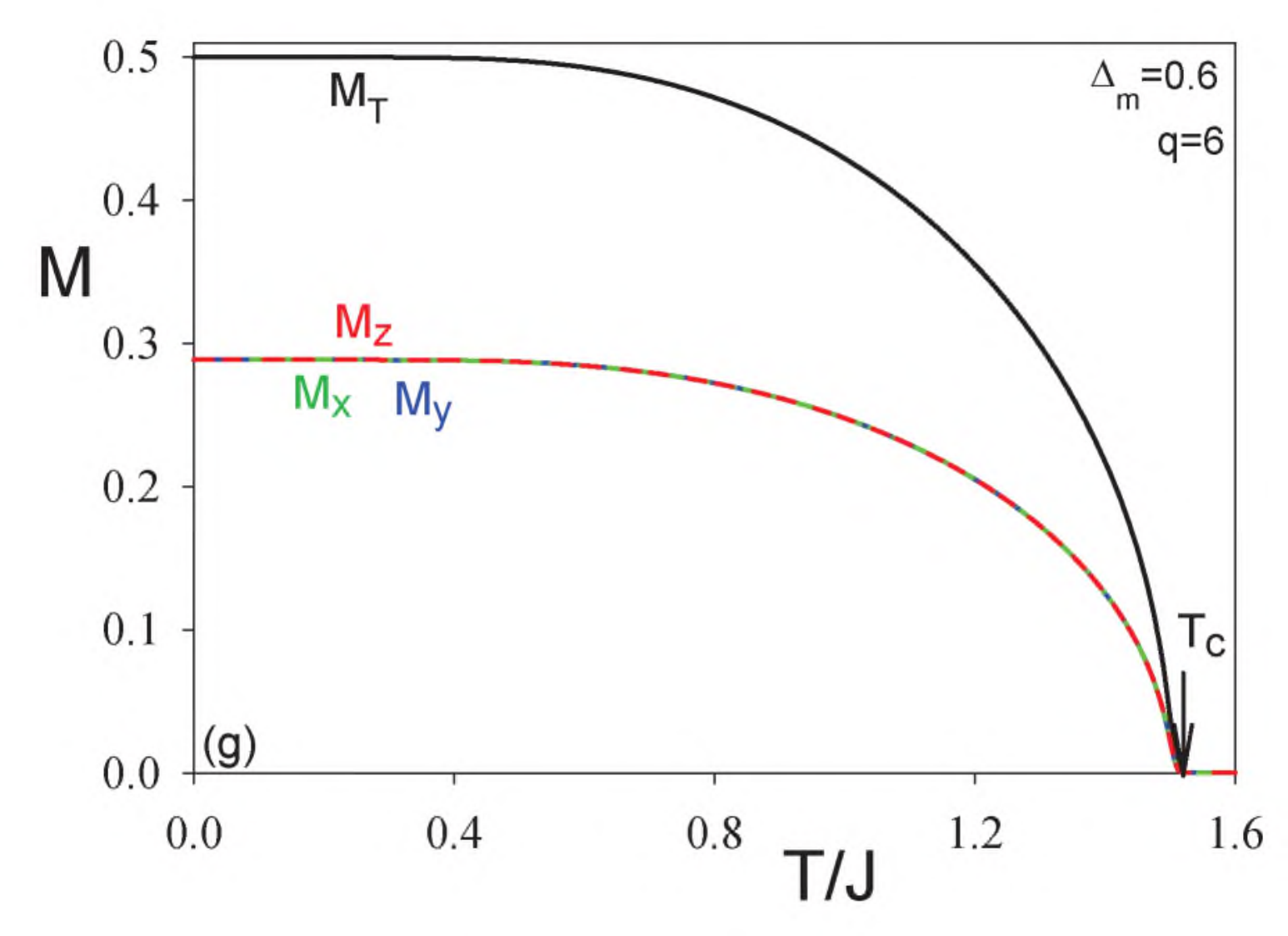}
	\end{center}
	
	\caption{(Colour online) The thermal variations of magnetization components and magnetization when $H=0$ and $q=6$ for the given values of $\Delta_m$ as 
		(a) $\Delta_{x}=\Delta_{y}=\Delta_{z} =0$ and $J_{x}=J_{y}=J_{z}$ case with $H=0$ and in the inset $H=1$,	(b) 1.3, (c) 1.2, (d) 1, (e) 0.94, (f) 0.9 and (g) 0.6.}
	\label{fig:1}
\end{figure}

\section{Thermal variations of magnetizations}
In this section, the characteristic thermal variations of magnetizations are illustrated when the external magnetic field $H$ is turned off and on. They are presented for $q=6$ only, because of the qualitative similarities with the $q=3$ and 4. There is only quantitative difference between them, i.e., the critical temperatures are observed at higher values for higher $q$ which is expected.

First, we perform a numerical calculation test for the simplified case of $\Delta_{x}=\Delta_{y}=\Delta_{z} =0$ and $J_{x}=J_{y}=J_{z}$ when the Hamiltonian~(\ref{eq:Hamiltonian}) reduces to the isotropic Heisenberg model in which the MFA method is well-established and can be used as reference. Figure~\ref{fig:1}~a shows the case with zero external magnetic field with all the magnetizations tending to zero at the same $T_c$ and the inset is obtained for $H=1$ shows that magnetizations do not tend to zero. These are the well-known results of the isotropic Heisenberg model. It should be noted that the behavior of magnetizations changes for the values of $\Delta_m=\Delta_x=\Delta_y=\Delta_z$ being large (I), greater than 1 but close to it (II), equal to one (III) and less than~1~(IV) in some characteristic forms for $H=0$ as we shall see later. Figure~\ref{fig:1}~b is obtained for the case~(I) with $\Delta_m=1.3$ and shows that magnetizations are less than 0.5 at zero temperature, and they decrease as the temperature decreases and eventually terminate at the second-order phase transition temperature ($T_c$) separating the FM and PM phases. It is also clear that $M_T>M_x=M_z=-M_y$. Case (II) is presented in figure~\ref{fig:1}~c for $\Delta_m=1.2$ and shows that magnetizations first exhibit some random behaviors moving up and down which terminates at the critical temperature called $T_R$, then their behaviors become similar to the case (I) with all the lines terminating at the $T_c$. When $\Delta_m$ is set equal to 1 for the case (III) as shown in figure~\ref{fig:1}~d, the magnetizations show only random behaviors from the beginning to the end. Now, $T_R$ separates the random phase region from the PM phase. Figure~\ref{fig:1}~e shows some branching of magnetizations calculated for $\Delta_m=0.94$. The lines start with random behaviors in the interval of some values corresponding to $M_T>M_x=M_z>M_y$ with branching. The random behavior terminates at the $T_R$, then the lines become regular curves but branching continues which terminates at a temperature called $T_B$. At $T_B$, we see that $M_x=M_y=M_z$. Then, they follow each other terminating at the $T_c$ as~$M_T$.  The branching  is still seen for $\Delta_m=0.9$ as shown in figure~\ref{fig:1}~f with no random behaviors anymore. Again, the branching terminates at $T_B$, then the rest is as in figure~\ref{fig:1}~e. Finally, for $\Delta_m=0.6$, all the components of magnetizations follow the same curve, i.e., $M_T>M_x=M_y=M_z$ and terminates at the~$T_c$. It is interesting to note that when the random behavior or branching appears, $M_T$ may be a little larger than~0.5, otherwise it is either smaller or equal to 0.5. It may be interesting to further search the reason of these fluctuations, and it is obvious that it has quantum mechanical origin. It should also be mentioned that these random behaviors in magnetizations must be caused by the existence of the DM interactions leading to the skyrmions which will not be examined in this work. 

\begin{figure}
	\begin{center}
		\includegraphics[width=6.9cm]{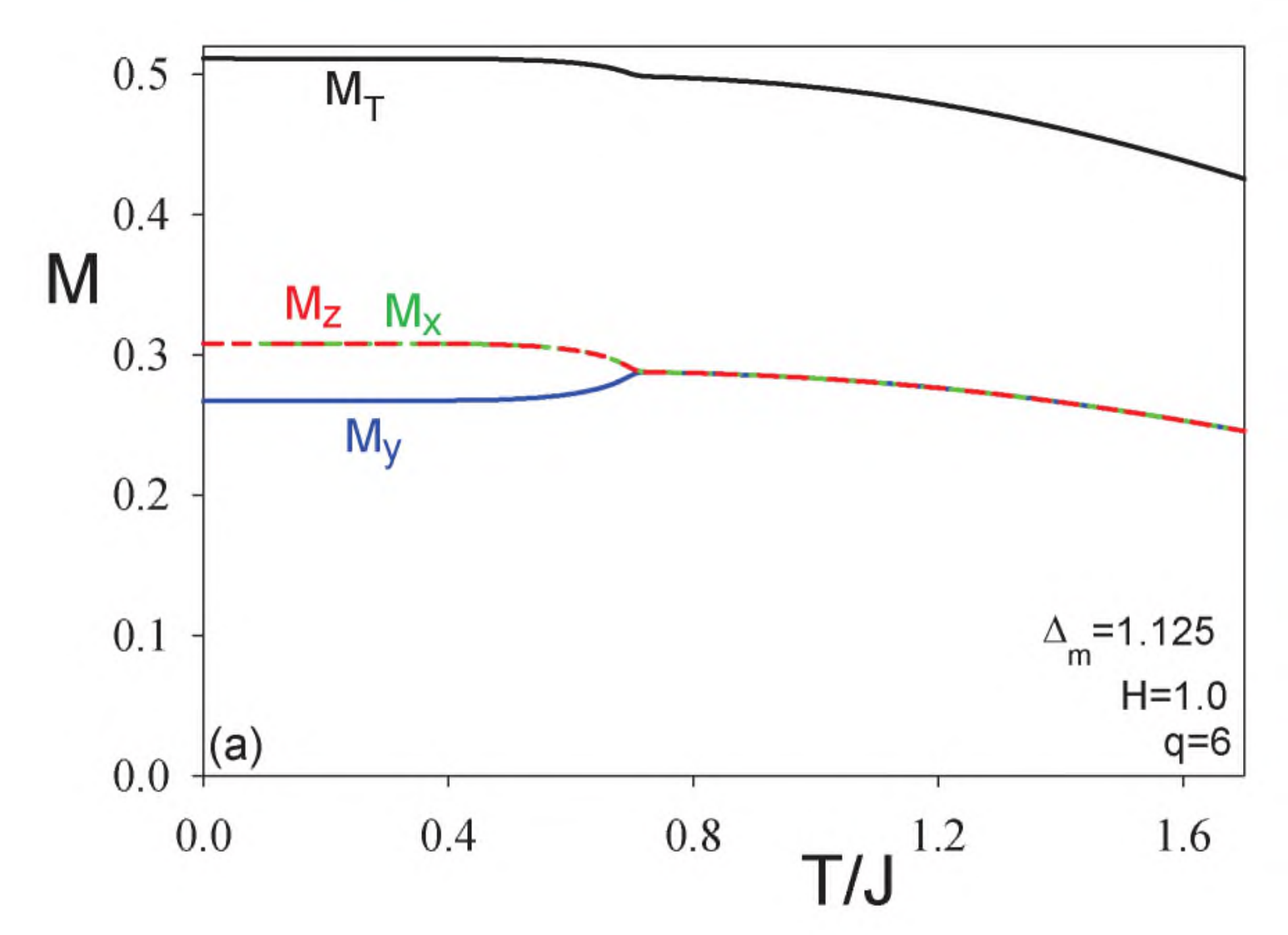}
		\includegraphics[width=6.9cm]{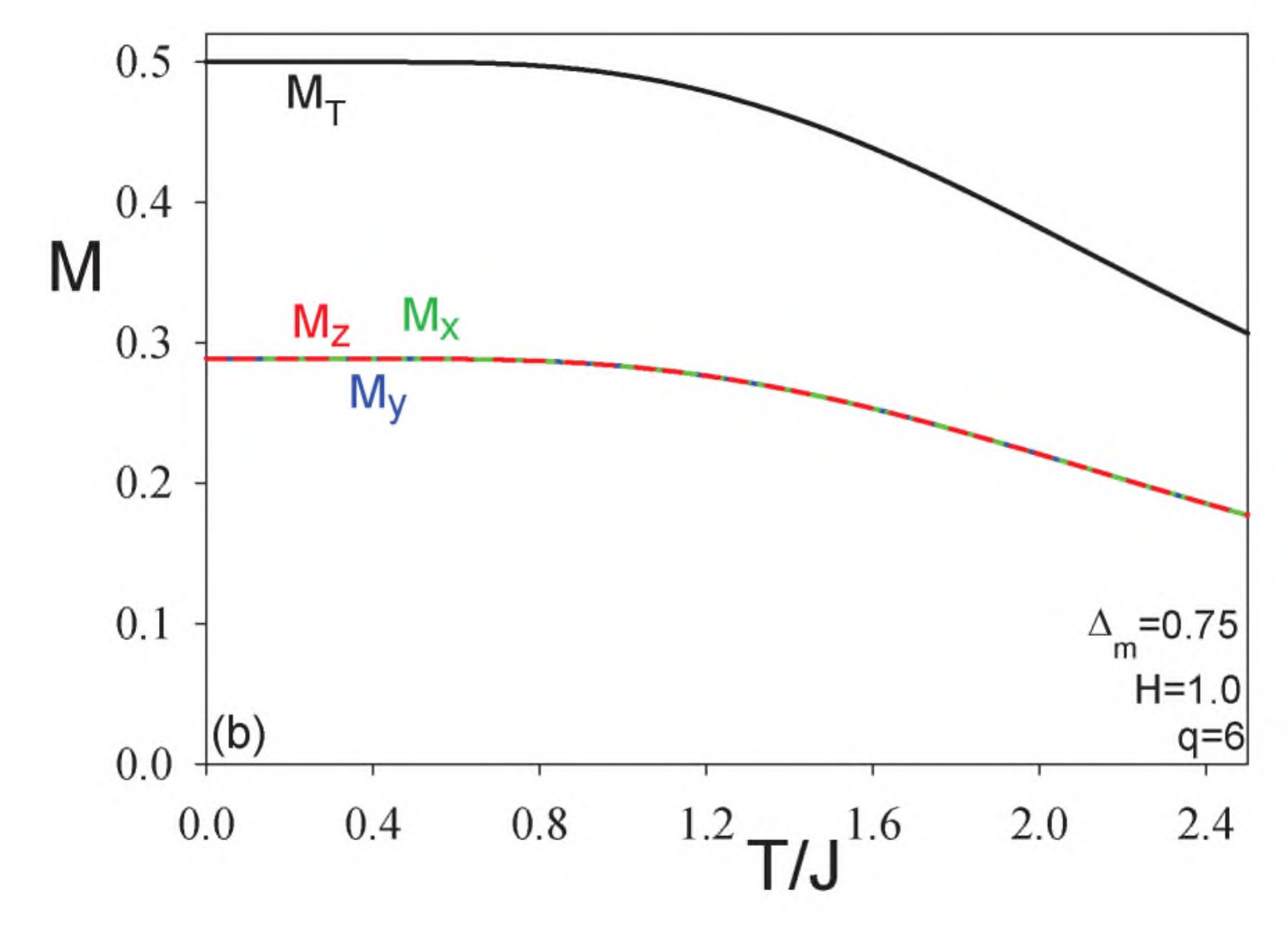}
	\end{center}
	\caption{(Colour online) Thermal variations of magnetization components and magnetization when $H=1$ and $q=6$ for the given values of $\Delta_m$ as 
		(a) 1.125 and (b) 0.75.}
	\label{fig:2}
\end{figure}

Thermal behaviors of magnetizations when $H$ is turned on shows two characteristic forms. When $\Delta_m>1$ and $H=1$, we see that branching continues as $M_x=M_z>M_y$ which terminates at $T_B$ where $M_x=M_y=M_z$ (see figure~\ref{fig:2}~a). Afterwards, they follow each other, never tending to the zero which is expected when $H$ is on. It is also clear that $M_T$ is a little larger than 0.5. When $\Delta_m \leqslant 1$ and $H=1$, we always see $M_T>M_x=M_y=M_z$ as shown in figure~\ref{fig:2}~b. Again, they do not tend to zero with increasing temperature.

In the next section, we combine all these critical temperatures to construct the phase diagrams on the ($\Delta_m/J, T/J$) planes when $H$ is turned off. The combinations of these points make up the possible phase lines and their combinations lead to some critical points.
  
\section{The phase diagrams}
Now, we construct the phase diagrams on the $(\Delta_m/J, T/J)$ planes for the given values of the coordination numbers $q=3, 4$ and 6. In the phase diagrams we identify five different phase regions:

\begin{itemize}
\item
The FM phase with constant $T_c$-line corresponding to the values of $\Delta_m \leqslant 1$ where $M_x=M_y=M_z$ as seen in figure~\ref{fig:1}~f (see the part after $T_B$) and in figure~\ref{fig:1}~g.
\item
The FM phase with linearly increasing $T_c$-line corresponding to the values of $\Delta_m>1$ where $M_x=M_z=-M_y$ as seen in figure~\ref{fig:1}~b and c after $T_R$ with the slopes of about 1.5, 2.0 and 3.0 for $q=3, 4$ and 6, respectively. It should be noted that these values are also the same values in the Ising model where the FM phase finishes and the PM phase starts, i.e., order-disorder phase transition temperatures, for $q=3, 4$ and 6, respectively. 
\item
The random phase regions indicated with (R) where magnetizations behave randomly going up and down as seen in figure~\ref{fig:1}~c--e before $T_R$. It is also interseting to see the branching of magnetization components in this phase.
\item
The FM phase region with branching indicated with (B) corresponding to $M_x=M_z\neq M_y$ as indicated in figure~\ref{fig:1}~e and f.
\item
The PM phase region with zero magnetizations caused by the thermal agitations. \\
The border lines between these phase regions, i.e., phase transition lines, are indicated with solid, dotted-dashed and dashed lines indicating the $T_c, T_R$ and $T_B$-lines, respectively.
\end{itemize}

\begin{figure}[htb]
\centerline{\includegraphics[width=0.45\textwidth]{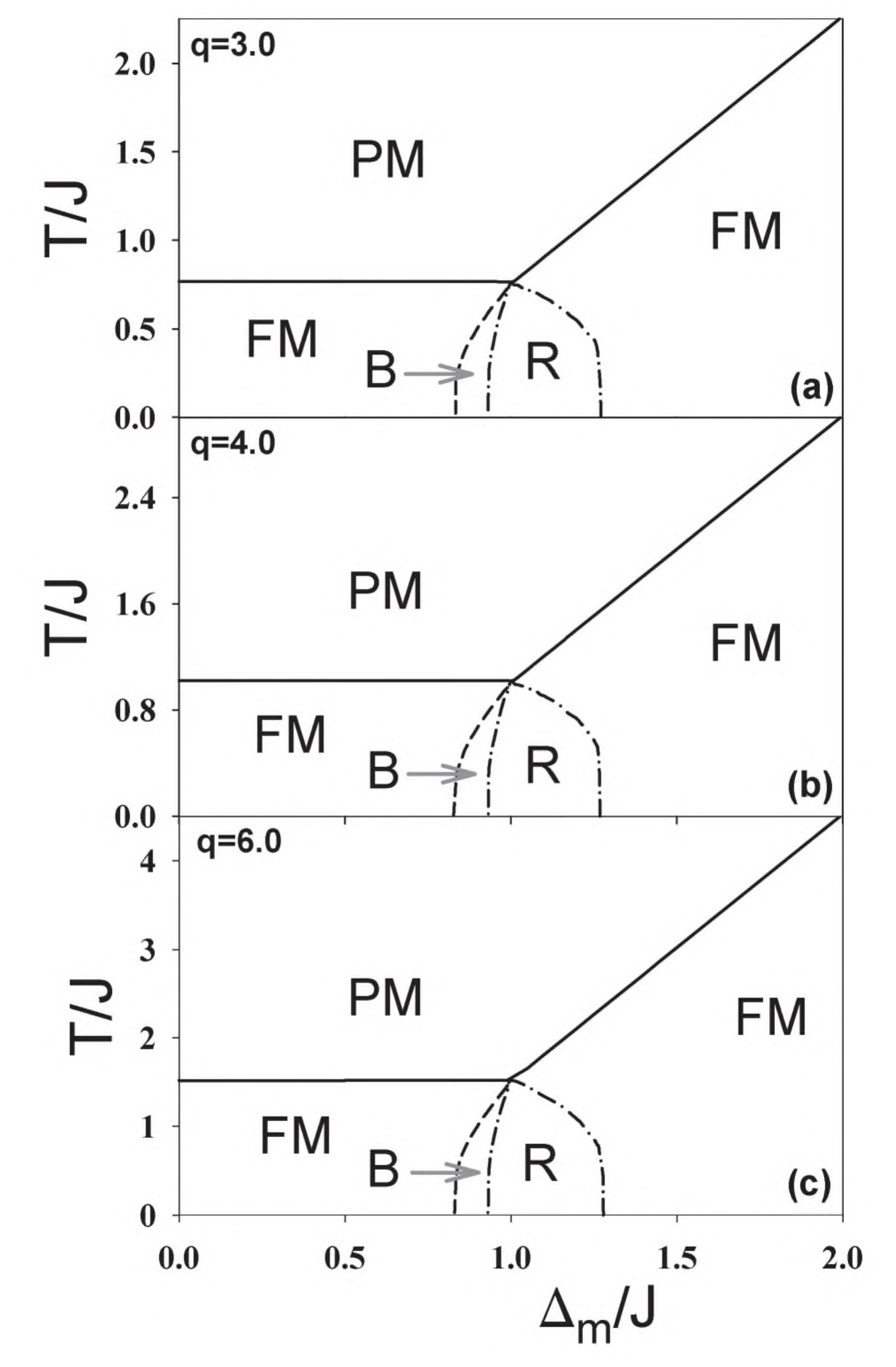}}
\caption{The phase diagrams on the $(\Delta_m/J, T/J)$ planes for the isotropic FM phase when $H$ is turned off for (a) $q=3.0$, (b) $q=4.0$ and (c) $q=6.0$.} \label{fig-3}
\end{figure}

As seen in figure~\ref{fig-3}, the phase diagrams are similar for all values of the coordination numbers. The $T_c$-lines consist of a straight part and a linearly increasing part. The straight parts are found at temperatures $T_c=0.7675, 1.02$ and 1.52 for $q=3, 4$ and 6, respectively. The $T_R$- and $T_B$-lines originate from 1 and are located around it. The two portions of the $T_R$-lines starting from 1 terminate at zero temperature for about $\Delta_m=0.97$ and $\Delta_m=1.27$ making a closed loop enclosing the R phase for all $q$, but with higher temperatures for higher $q$. The $T_B$-lines terminate at the same value of $\Delta_m$ being about 0.83 for all $q$. The $T_B$-lines separate FM and R phases when $\Delta_m<1$. The $T_R$-line when $\Delta_m>1$ separates the FM phase from the R phase. It is clear that the model does not produce any first-order phase transition lines as in the well-known Ising model. It should also be noted that the $\Delta_m=1$ is a special critical point from where two $T_c$-, two $T_R$- and one $T_B$-lines emerge. 

\section{Summary and conclusions}
%\hspace{0.5cm} 
The spin-1/2 Heisenberg Hamiltonian for the FM case with $J=J_x=J_y=J_z>0$ is considered to study the effects of DM interactions with $\Delta_m=\Delta_x=\Delta_y=\Delta_z$ on the thermal variations of the magnetization components and magnetization when the external magnetic field components $H=H_x=H_y=H_z$ are turned off and on. The phase diagrams of the model are calculated on the ($\Delta_m/J, T/J$) planes for $q=3, 4$ and 6. Three different FM phase regions are observed with $M_x=M_y=M_z$, $M_x=M_z=-M_y$ and the one with $M_x=M_z \neq M_y$ exhibits branching in addition to the PM and random phase regions. It should be noted that this model has not been studied before, so the comparison is not possible. As a last word, the anisotropic case of this model is going to be considered as a continuation of this work. 

\section*{Acknowledgements}
This work was supported by the Research Fund of Erciyes University with Project Identification Number: FBA-2021-11571.

\ukrainianpart

\title{Феромагнітна модель Гайзенберга зі взаємодією Дзялошинського-Морія%
}
\author{E. Албайрак}

\address{Фізичний факультет, Університет Ерчиєс, 38039 Кайсері, Туреччина
}

\makeukrtitle
\begin{abstract}
	Спін-1/2 модель Гайзенберга сформульовано в наближенні середнього поля з використанням матричної форми спінових операторів  $\hat{S}_x,\hat{S}_y$ і $\hat{S}_z$ у тривимірному просторі. У гамільтоніан, що розглядається, входять параметри білійнійної обмінної взаємодії $(J_x,J_y,J_z)$, взаємодії Дзялошинського-Морія $(\Delta_x,\Delta_y,\Delta_z)$, а також компоненти зовнішнього магнітного поля $(H_x,H_y,H_z)$. Компоненти намагніченості отримані у наближенні середнього поля для загального анізотропного випадку, коли $J_x\neq J_y \neq J_z$, для різних координаційних чисел $q$. Температурні залежності намагніченостей досліджено детально з метою побудови фазових діаграм моделі для ізотропного випадку $J_x=J_y=J_z>0$. Виявлено, що для моделі існують феромагнітна, парамагнітна, випадкова фазові області, а також додаткова феромагнітна фаза, в якій у компонент намагніченостей спостерігається галуження.
	
	
\end{abstract}

\lastpage

\begin{thebibliography}{99}
%1
	\bibitem{DM} Dzyaloshinsky I., J. Phys. Chem. Solids, 1958, \textbf{4}, 241, \doi{10.1016/0022-3697(58)90076-3}.
%2
	\bibitem{DM2} Moriya T., Phys. Rev. Lett., 1960, \textbf{4}, 228, \doi{10.1103/PhysRevLett.4.228}.
%3		
	\bibitem{Komatsu} Komatsu H., Nonomura Y., Nishino M., Phys. Rev. B,  2021, \textbf{103}, 214404, \doi{10.1103/PhysRevB.103.214404}. 
%4
	\bibitem{Nishikawa} Nishikawa Y., Hukushima K., Phys. Rev. B, 2016, \textbf{94}, 064428, \doi{10.1103/PhysRevB.94.064428}.
%5	
	\bibitem{Shinozaki} Shinozaki M., Hoshino S., Masaki Y., Kishine J., Kato Y., J. Phys. Soc. Jpn., 2016, \textbf{85}, 074710,\\ \doi{10.7566/JPSJ.85.074710}.	
%6	
	\bibitem{ExpSky} M\"{u}hlbauer S., Binz B., Jonietz F., Pfleiderer C., Rosch A., Neubauer A., Georgii R., B\"{o}ni P., Science, 2009, \textbf{323}, 915, \doi{10.1126/science.1166767}.
%7
	\bibitem{ExpSky2} Yu X. Z., Onose Y., Kanazawa N., Park J. H., Han~J.~H., Matsui~Y., Nagaosa~N., Tokura~Y., Nature, 2010, \textbf{465}, 901, \doi{10.1038/nature09124}.
%8	
	\bibitem{ExpSky3} Jonietz F., M\"{u}hlbauer S., Pfleiderer C., Neubauer A., M\"{u}nzer W., Bauer A., Adams T., Georgii R., B\"{o}ni P., Duine R. A., Everschor K., Garst M., Rosch A., Science, 2010, \textbf{330}, 1648, \doi{10.1126/science.1195709}.
%9	
	\bibitem{ExpSky4} Yu X. Z., Kanazawa N., Onose Y., Kimoto~K., Zhang~W.~Z., Ishiwata~S., Matsui~Y., Tokura~Y., Nat. Mater., 2011, \textbf{10}, 106, \doi{10.1038/nmat2916}.
%10	
	\bibitem{ExpSky5} Schulz T., Ritz R., Bauer~A., Halder~M., Wagner~M., Franz~C., Pfleiderer~C., Everschor~K., Garst~M., Rosch~A., Nat. Phys., 2012, \textbf{8}, 301, \doi{10.1038/nphys2231}.
%11	
	\bibitem{ExpSky6} Sampaio J., Cros V., Rohart S., Thiaville A., Fert A., Nat. Nanotechnol., 2013, \textbf{8}, 839,\\ \doi{10.1038/nnano.2013.210}.
%12	
	\bibitem{ExpSky7} Jiang W., Upadhyaya P., Zhang W., Yu G., Jungfleisch M. B., Fradin F. Y., Pearson~J.~E., Tserkovnyak~Y., Wang~K.~L., Heinonen~O., te~Velthuis~S.~G.~E., Hoffmann~A., Science, 2015, \textbf{349}, 283,\\ \doi{10.1126/science.aaa1442}.
%13	
	\bibitem{ExpSky8} Boulle O., Vogel J., Yang~H., Pizzini~S., de~Souza~Chaves~D., Locatelli~A., Mentes~T.~O., Sala~A., \mbox{Buda-Prejbeanu}~L.~D., Klein~O., Belmeguenai M., Roussign\'{e} Y., Stashkevich A., Ch\'{e}rif S. M., Aballe~L., Foerster M., Chshiev M., Auffret S., Miron I. M., Gaudin G., Nat. Nanotechnol., 2016, \textbf{11}, 449,\\ \doi{10.1038/nnano.2015.315}.
%14	
	\bibitem{ExpSG} Pr\'ejean J. J., Joliclerk M. J., Monod P., J. Phys. (Paris), 1980, \textbf{41}, 427, \doi{10.1051/jphys:01980004105042700}. 
%15	
	\bibitem{ExpSG2} Hippert F., Alloul H., Pr\'ejean J. J., Physica B, 1981, \textbf{107}, 427. 
%16	
	\bibitem{TheoSG} Levi P. M., Fert A., Phys. Rev. B, 1981, \textbf{23}, 4667, \doi{10.1103/PhysRevB.23.4667}. 
%17	
	\bibitem{TheoSG2} Levi P. M., Morgan-Pond C., Fert A., J. Appl. Phys., 1982, \textbf{53}, 2168, \doi{10.1063/1.330770}.
%18
	\bibitem{BLi} Li B., Cho S. Y., Wang H. L., Hu B. Q., J. Phys. A: Math. Theor., 2011, \textbf{44}, 392002, \doi{10.1088/1751-8113/44/39/392002}.
%19	
	\bibitem{Soltanix} Soltani M. R., Mahdavifar S., Akbari A., Masoudi A. A., J. Supercond. Novel Magn., 2010, \textbf{23}, 1369,\\ \doi{10.1007/s10948-010-0785-x}.
%20	
	\bibitem{Strecka} Stre\v{c}ka J., \v{C}anov\'{a} L., J. Phys.: Conf. Ser., 2009, \textbf{145}, 012012, \doi{10.1088/1742-6596/145/1/012012}.
%21	
	\bibitem{Parente} Parente W. E. F., Pacobahyba J. T. M., Ar\'{u}jo I. G., Neto M. A., de Sousa J. R., Akinci \"{U}., 	J. Magn. Magn. Mater., 2014, \textbf{355}, 235, \doi{10.1016/j.jmmm.2013.12.041}.
%22	
	\bibitem{Liu1} Liu G. H., You W. L., Li W., Su G.,  J. Phys.: Condens. Matter, 2015, \textbf{27}, 165602, \doi{10.1088/0953-8984/27/16/165602}.
%23	
	\bibitem{Strecka1} Stre\v{c}ka J., \v{C}anov\'{a} L., Minami K., Phys. Rev. E, 2009, \textbf{79}, 051103, \doi{10.1103/PhysRevE.79.051103}.
%24	
	\bibitem{Parente1} Parente W. E. F., Pacobahyba J. T. M., Neto M. A., Ara\'{u}jo I. G., Plascak J. A., J. Magn. Magn. Mater., 2018, \textbf{462}, 8, \doi{10.1016/j.jmmm.2018.04.054}.
%25	
	\bibitem{Avalishvili} Avalishvili N., Beradze B., Japaridze G. I., Eur. Phys. J. B, 2019, \textbf{92}, 262, \doi{10.1140/epjb/e2019-100323-1}. 
%26	
	\bibitem{Chan} Chan Y. H., Jin W., Jiang H. C., Starykh O. A., Phys. Rev. B, 2017, \textbf{96}, 214441,\\ \doi{10.1103/PhysRevB.96.214441}.
%27	
	\bibitem{Flynn} Flynn M. O., Singh R. R. P., Phys. Rev. B, 2019, \textbf{100}, 121108(R), \doi{10.1103/PhysRevB.100.121108}.
%28	
	\bibitem{Freitas} Freitas A. S., de Albuquerque D. F., Phys. Rev. E, 2015, \textbf{91}, 012117, \doi{10.1103/PhysRevE.91.012117}.
%29	
	\bibitem{Grandi} Grandi N., Lagos M., Oliva J., Vera A., Eur. Phys. J. B, 2019, \textbf{92}, 244, \doi{10.1140/epjb/e2019-100395-3}.
%30	
	\bibitem{Griset} Griset C., Head S., Alicea J., Starykh O. A., Phys. Rev. B, 2011, \textbf{84}, 245108, \doi{10.1103/PhysRevB.84.245108}.
%31	
	\bibitem{Japaridze} Japaridze G. I., Cheraghi H., Mahdavifar S., Phys. Rev. E, 2021, \textbf{104}, 014134, \doi{10.1103/PhysRevE.104.014134}.
%32	
	\bibitem{XLi} Li X., Jin J., Phys. Rev. B, 2021, \textbf{103}, 035127, \doi{10.1103/PhysRevB.103.035127}.
%33	
	\bibitem{Messio} Messio L., Bieri S., Lhuillier C., Bernu B., Phys. Rev. Lett., 2017, \textbf{118}, 267201,\\ \doi{10.1103/PhysRevLett.118.267201}.
%34	
	\bibitem{Sera} Sera A., Kousaka Y., Akimitsu J., Sera M., Kawamata T., Koike Y., Inoue K., Phys. Rev. B, 2016, \textbf{94}, 214408,\\ \doi{10.1103/PhysRevB.94.214408}.
%35	
	\bibitem{Shindou} Shindou R., Phys. Rev. B, 2016, \textbf{93}, 094419, \doi{10.1103/PhysRevB.93.094419}.
%36	
	\bibitem{Szymczak} Szymczak H., Baran M., Szymczak R., Barilo S. N., Bychkov G. L., Shiryaev S. V., Acta Phys. Pol., A, 2007, \textbf{111}, 71, \doi{10.12693/APhysPolA.111.71}
	
\end{thebibliography}
\end{document}